\documentclass[aps,prl,twocolumn,showpacs,superscriptaddress]{revtex4}
\pdfoutput=1
\usepackage{graphics,graphicx}
\usepackage{color}
\usepackage{wrapfig}
\usepackage{enumerate}
\usepackage{amssymb,amsmath}
\usepackage{bm}

\begin{document}
\unitlength = 1mm
%~~~~~~~~~~~~~~~~~~~~~~~~~~~~~~~~~~~~~~~~~~~~~~~~~
\title{
First-principles design of the spinel iridate Ir$_2$O$_4$ for high-temperature quantum spin ice
}

\author{Shigeki Onoda}
\affiliation{Condensed Matter Theory Laboratory, RIKEN, Wako 351-0198, Japan}
\affiliation{Quantum Metter Theory Research Team, RIKEN Center for Emergent Matter Science (CEMS), Wako 351-0198, Japan}
\author{Fumiyuki Ishii}
\affiliation{Faculty of Mathematics and Physics, Institute of Science and Engineering, Kanazawa University, Kanazawa 920-1192, Japan}

\date{\today}
%75.30.Kz Magnetic phase transition
%
%%%%%%%%%%% %%%%%%%%%%% %%%%%%%%%%% %%%%%%%%%%% %%%%%%%%%%% %%%%%%%%%%%
\begin{abstract}
  Insulating magnetic rare-earth pyrochlores related to spin ice host emergent bosonic monopolar spinons, which obey a magnetic analogue of quantum electrodynamics and may open a route to a magnetic analogue of electronics.
  However, the energy scales of the interactions among rare-earth moments are so low as 1~K that the possible quantum coherence can only be achieved at a sub-Kelvin. Here, we desgin high-temperature quantum spin ice materials from first principles. It is shown that the $A$-site deintercalated spinel iridate Ir$_2$O$_4$, which has been experimentally grown as epitaxial thin films, is a promising candidate for quantum spin ice with a spin-ice-rule interaction of a few tens of meV. Controlling electronic structures of Ir$_2$O$_4$ through substrates, it is possible to tune magnetic interactions so that a magnetic Coulomb liquid persists at high temperatures.
\end{abstract}

\maketitle

A qust to quantum spin liquids (QSLs), showing no long-range magnetic order but hoscting fractionalized quasiparticles, dubbed spinons, has been one of central issues in condensed-matter physics~\cite{lee:08,balents:10}. Among many kinds of QSLs, quantum spin ice (QSI) systems~\cite{molavian:07,onoda:09} offer a unique laboratory for hosting bosonic spinons, dubbed QSI monopoles, carrying hedge-hog monopole charges on the dual diamond lattice site~\cite{hermele:04,moessner:06} (Fig.~1(a)). The ideal ground state is a bosonic U(1) QSL~\cite{hermele:04}, which forms a magnetic analogue of an ``electrical insulator'' in quantum electrodynamics (QED): QSI monopoles interact through a magnetic Coulomb interaction mediated by gapless analogous ``photon'' modes in spin excitations~\cite{hermele:04} and possess a finite kinetic energy but remain incompressible because of an energy gap in the excitation spectrum. This may open novel potential applications along the concept of a magnetic analogue of electronics, where QSI monopoles take place of electrons.
QSI systems are found in magnetic rare-earth pyrochlore oxides~\cite{gardner:10}, including Yb$_2$Ti$_2$O$_7$~\cite{chang:13,ross:11}, Tb$_2$Ti$_2$O$_7$~\cite{takatsu:15}, and Pr$_2$Zr$_2$O$_7$~\cite{kimura:13}, that are variants of spin ice~\cite{harris:97,bramwell:01}. However, the temperature/energy scale of the interaction beween the nearest-neighbor magnetic moments is too low ($\sim1$~K) to observe the quantum coherence. QSI materials that accommodate the analogous QED in a much higher temperature scale are called for.

In order to increase the temperature scale, effective spin-$\frac{1}{2}$ magnetic moments of correlated transition-metal $d$ electrons localized by an on-site Coulomb repulsion $U$ are advantageous, since the wavefunction overlap and thus the exchange couplings between the magnetic moments become much larger than for the rare-earth $4f$ electrons. Simultaneously, a large magnetic anisotropy in the exchange interactions and thus a large relativistic spin-orbit coupling are indispensable for the QSI physics. These considerations highlight strongly spin-orbit coupled Kramers doublets at half filling, for instance, based on Ir$^{4+}$ ions under the predominant cubic crystalline electric field~\cite{kim:09}.
There is another severe constraint: interations among spins on the pyrochlore lattice should favour two spins pointing inwards to and the other two outwards from the center in each tetrahedron. (See Fig.~1(a).) To accommodate this ferromagnetic ``2-in, 2-out'' spin-ice-rule interaction~\cite{harris:97,bramwell:01}, the bond angle formed by a nearest-neighbor Ir-O-Ir bond angle should be close to 90$^\circ$, according to the Kanamori-Goodenough rule. This is not the case in pyrochlore iridates $A_2$Ir$_2$O$_7$, where the ``all-in, all-out'' antiferromagnetism (Fig.~1(a)) is stablized in the strong-coupling case~\cite{wan:11,ishii:15,tomiyasu:12,sagayama:13}.
On the other hand, in spinel compounds $A$Ir$_2$O$_4$ where Ir sites also form the pyrochlore lattice structure, the bond angle is much closer to 90$^\circ$ (Fig.~1(b)), potentially hosting the spin-ice-rule interaction.
Remarkably, it has been reported that the $A$-site deintercalated spinel iridate Ir$_2$O$_4$ with the Ir$^{4+}$ ions can be obtained from epitaxial thin films of Li$_x$Ir$_2$O$_4$ at least on the LiNbO$_3$(0001) substrate~\cite{kuriyama:10}. Transport and optical measurements indicated that it is a narrow-gap (34~meV) Mott insulator~\cite{kuriyama:10} without showing a clear anomaly for a magnetic phase transition. However, detailed magnetic properties of the material remain open. 

In this Letter, we provide evidence from first principles that the $A$-site de-intercalated spinel iridate Ir$_2$O$_4$ hosts a dominant spin-ice rule interaction of the order of room temperature as well as additional subdominant magnetic interactions. We tune the magnetic interactions by a tetragonal distortion through the substrate, to design high-temperature quantum spin ice.
\begin{table}
  \caption{\label{table:structure}Crystal parameters of Ir$_2$O$_4$ determined from fully relativistic LSDA structure stability calculations.}
  \begin{tabular}{llll}
    \\
    \hline\hline 
    \multicolumn{4}{c}{$Fd\bar{3}m$: $a_{\mathrm{c}}=8.7119$~\AA}
    \\ \hline
    Site & $f_a$  & $f_b$  & $f_c$
    \\
    Ir ($16d$) & 0.5 & 0.5 & 0.5
    \\ 
    O ($32e$)  & $x$=0.26911 & $x$ & $x$
    \\ \hline\hline
    \multicolumn{4}{c}{$I4_1/amd$: $a_{\mathrm{t}}=\sqrt{2}a_{\mathrm{MgO}}=5.955$~{\AA}~\cite{kuriyama:10}, $c_{\mathrm{t}}=8.748$~\AA}
    \\ \hline
    Ir ($8d$) &  0 &  0 & 0.5
    \\
    O ($16h$) &  0  & 0.15754 & 0.26713
    \\ \hline\hline
    \multicolumn{4}{c}{ $R\bar{3}m$: $a_{\mathrm{r}}=6.021$~{\AA}~\cite{kuriyama:10}, $c_{\mathrm{r}}=15.346$~\AA}
    \\ \hline
    Ir ($3a$) & 0 & 0 & 0
    \\ 
    Ir ($9d$) & 0.5 & 0.5 & 0.5
    \\ 
    O  ($6c$) & 0 & 0 & 0.2332
    \\ 
    O  ($18h$)& $x$=0.5232 & $-x$ & $z$=0.2449
    \\ \hline\hline
  \end{tabular}
\end{table}

\begin{figure}
\begin{center}
\includegraphics[width=8.0cm]{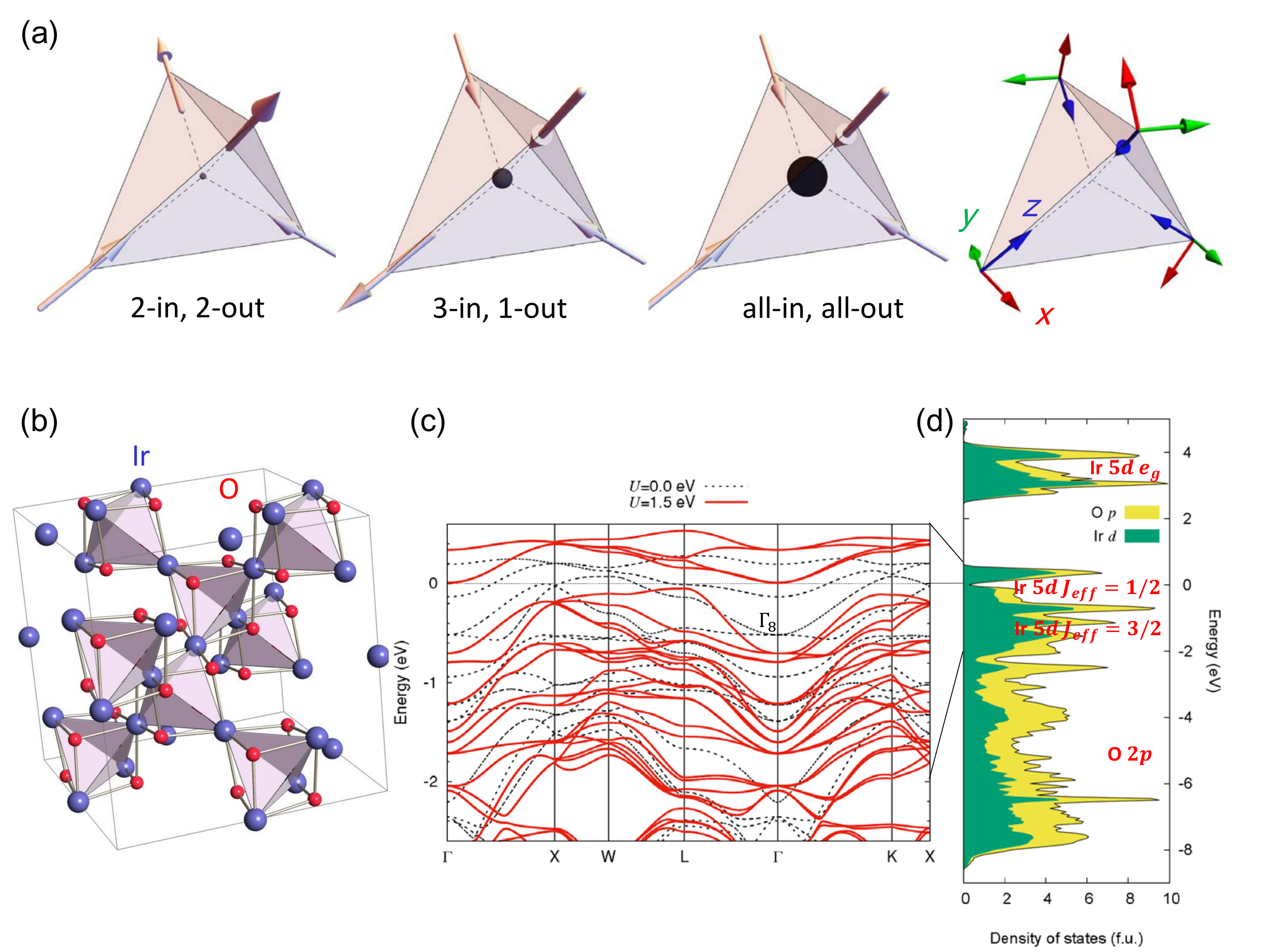}
\caption{(Color online) Cystal and electronic structures of the $A$-site deintercalated cubic spinel iridate Ir$_2$O$_4$.
  (a), From left to right, ``2-in, 2-out'', ``3-in, 1-out'' and ``all-in, all-out'' spin configurations hosting 0, +1 and +2 monopole charges, respectively, and the sublattice-dependent $C_2$-invariant local coordinate frames $U_{bm{r}_\mu}=(\bm{x}_\mu,\bm{y}_\mu,\bm{z}_\mu)$~\cite{note:LocalFrames}.
  (b), Crystal structure of Ir$_2$O$_4$.
  (c), Paramagnetic ($Fd\bar{3}m'$) and antiferromagnetic ($I4_1'/am'd$) electronic band structures of Ir$_2$O$_4$ are shown with black dashed curves and red solid curves, respectively. They have been obtained with the LSDA+$U$ calculations with $U=0$~eV and $U=1.5$~eV.
  (d), Density of states of electrons for the antiferromagnetic ($I4_1'/am'd$) state.}
\label{fig:structure}
\end{center}
\end{figure}

Let us start with the hypothetical cubic $A$-site-deintercalated spinel iridate Ir$_2$O$_4$ as a reference. Figure~1(b) as well as Table~1 shows its ideal crystal structure obtained by relativistic local spin density approximation (LSDA) calculations~\cite{note:Openmx}. The paramagnetic electronic band structure is shown with dashed curves in Fig.~1(c). Across the Fermi level, it displays four Kramers pairs of band dispersions that are disentangled from the others as in Pr$_2$Ir$_2$O$_7$~\cite{ishii:15}, allowing for an interpretation that each Ir ion offers a single Kramers doublet sharing common properties with $J_{\mathrm{eff}}$-$\frac{1}{2}$ spinors~\cite{kim:09,witczak-krempa:13}. Striking differences from Pr$_2$Ir$_2$O$_7$~\cite{ishii:15} are in order: (i) the bandwidth of this $J_{\mathrm{eff}}$-$\frac{1}{2}$ manifold is reduced by 10-20~$\%$ and (ii) the $\Gamma_8$ quartet marked in Fig.~1(c), which is located near the Fermi level in Pr$_2$Ir$_2$O$_7$, lies deeply below the Fermi level. We then include an on-site Coulomb interaction $U$ among Ir $d$ electrons by the LSDA+$U$ method~\cite{anisimov:91}. We find several (meta)stable translation-invariant solutions labeled by a magnetic space group $MSG$. The energy $E_{MSG}$ and nonzero ordered magnetic moments $\bm{m}_{MSG}^\mu$ at the Ir sites with the sublattice $\mu$ are shown as functions of $U$ for each solution in Fig.~2(a) and 2(b), respectively. Increasing $U$ beyond 0.6~eV induces a phase transition from a paramagnetic metal of $Fd\bar{3}m1'$ to a metallic [100] ``2-in, 2-out''-like splayed ferromagnet of $I4_1/am'd'$. A level crossing at $U\sim1.2$~eV seen in Fig.~2(a) points to the second phase transition to an insulating antiferromagnet of $I4'_1/am'd$ or $I4_1/amd$. The degeneracy of these two antiferromagnetic solutions is reminiscenet of Er$_2$Ti$_2$O$_7$, where the $I4'_1/am'd$ antiferromagnetism is eventually selected by quantum order by disorder~\cite{champion:03}. Metastable magnetically ordered solutions include in the ascending order in energy, the ``2-in, 2-out''-like splayed [100]-ferromagnet, a ``3-in, 1-out''-like [111]-ferromagnet of $R\bar{3}m'$ for $U>0.9$~eV and two antiferromagnets of $I4'_1/amd'$ and $Fd\bar{3}m'$ (``all-in, all-out'') for $U>1.4$~eV. Images of the magnetically ordering patterns of all the above magentic space groups are depicted in Fig.~3.
Remarkably, the ``all-in, all-out'' antiferromagnetic state ($Fd\bar{3}m'$) has a rather high energy, remain metallic for $U\le2.5$~eV and then becomes completely unstable for larger $U$, in sharp constrast to pyrochlore iridates $A_2$Ir$_2$O$_7$~\cite{wan:11,ishii:15,witczak-krempa:13}.
Most likely, a realistic value of $U$ is enhanced from the Pr$_2$Ir$_2$O$_7$ case ($U=1.3$~eV)~\cite{ishii:15} because of the narrower bandwith. A choice of $U=1.5$~eV  produces a direct charge gap of about 40~meV, as seen in the band structure shown in Fig.~1(c), in reasonable agreement with the experimental observation of the activation energy 34~meV in the electric resistivity in Ir$_2$O$_4$~\cite{kuriyama:10},  The density of states shown in Fig.~1(d) is also consistent with the picture obtained from the optical measurement~\cite{kuriyama:10}. 

The above five Mott-insulating solutions of $I4_1'/am'd$, $I4_1/amd$, $I4_1'/amd'$, $R\bar{3}m'$ and $I4_1/am'd'$ obtained from the cubic Ir$_2$O$_4$ can be modeled by the generic nearest-neighbor pyrochlore pseudospin-$\frac{1}{2}$ Hamiltonian for Kramers doublets~\cite{onoda:10},
\begin{eqnarray}
  H=&\frac{J}{2}\sum_{\langle\bm{r},\bm{r}'\rangle}\left[
    \tau^z_{\bm{r}}\tau^z_{\bm{r}'}
    +\delta \tau^+_{\bm{r}}\tau^-_{\bm{r}'}
    +qe^{2i\phi_{\bm{r},\bm{r}'}}\tau^+_{\bm{r}}\tau^+_{\bm{r}'}
    \right.\nonumber\\
    &\left.+Ke^{i\phi_{\bm{r},\bm{r}'}}(\tau^z_{\bm{r}}\tau^+_{\bm{r}'}+\tau^+_{\bm{r}}\tau^z_{\bm{r}'})
    \right]+h.c.,
  \label{eq:H}
\end{eqnarray}
where $\bm{\tau}_{\bm{r}}=(\tau_{\bm{r}}^x,\tau_{\bm{r}}^y,\tau_{\bm{r}}^z)$ is a pseudospin-$\frac{1}{2}$ operator representing the Kramers doublet at the Ir site $\bm{r}$ in the $C_2$-invariant set of local spin frames $U_{\bm{r}_\mu}=(\bm{x}_\mu,\bm{y}_\mu,\bm{z}_\mu)$ shown in Fig.~1(a) and described in detail in Methods, with the spin-ice-rule coupling $J$ and three dimensionless couplings $\delta$, $q$ and $K$ and the bond-dependent phase $\phi_{\bm{r},\bm{r}'}$ associated with a choice of $U_{\bm{r}_\mu}$~\cite{note:LocalFrames}. Roles of the three dimensionless couplings are in order~\cite{savary:12,lee:12}. The $\delta$ term stabilizes the U(1) QSLs, unless $\delta<\delta_c\sim-0.104$~\cite{banerjee:08} where it yields the antiferromagnetsim of $I4'_1/am'd$ or $I4_1/amd$. The $q$ term induces the $I4'_1/amd'$ antiferromagnetism if $|q|>q_c\sim0.15$. The $K$ term destabilizes the U(1) QSL, eventually leading to the splayed ferromagnetism with a large rotation angle $\varphi$ of the ordered moments from the local $\langle111\rangle$ axess towards the global [100] direction, as in Yb$_2$Ti$_2$O$_7$~\cite{chang:13,yasui:03}.
It is found that $J=28$-$30$~meV and $(\delta,q,K)=(-0.38$-$0.40,-0.63$-$0.60, -0.05$-$0.07)$, depending on $U$, perfectly reproduce the energy differences and the spin structures of all the five solutions in the insulating region $U=2.6$-$3.0$~eV within the classical analysis~\cite{note:ExchangeCouplings}. The similar coupling constants can also explain the intermediate-$U$ region near $U=1.5$~eV of our choice. Hence, we conclude that Ir$_2$O$_4$ hosts a spin-ice-rule interaction of the order of room temperature, i.e., $30$~meV. This is two orders of magnitude larger than that in magnetic rare-earth pyrochlore oxides.
The small value of $K$ we obtained is advantageous for the U(1) QSL. This sharply contrasts to Yb$_2$Ti$_2$O$_7$ where a rather large $K$ value yields a nearly collinear [100] ferromagnet ($I4_1/am'd'$) with $\varphi\sim54.7^\circ$~\cite{chang:13,yasui:03}. On the other hand, the magnitudes of $\delta$ and $q$ need to be suppressed by about a factor of 4 from the current hypothetical cubic Ir$_2$O$_4$, in order to reach the U(1) QSL.

In reality, Ir$_2$O$_4$ compounds are grown as epitaxial thin films, so that one may be able to control the electronic properties through substrates. And the U(1) QSL ground state, if it emerges, should survive small but finite lattice distortions as far as analogous ``photon'' modes in spin excitations remain stable.  Then, a possible signature of the U(1) QSL in the first-principles approach is that (i) all the ``2-in, 2-out''-like ferromagnetic insulating solutions, whose energies can be slightly split off by the distortion, have lower energies than any other magnetically ordered insulating solutions and simultaneously, (ii) to ensure that $K$ is small enough, the rotation angle $\varphi$ in these ``2-in, 2-out''-like states is as small as $\varphi=24.0^\circ$ with $U=3.0$~eV or $27.3^\circ$ with $U=1.5$~eV for the above cubic Ir$_2$O$_4$. 

\begin{figure}
\begin{center}
\includegraphics[width=0.5\textwidth]{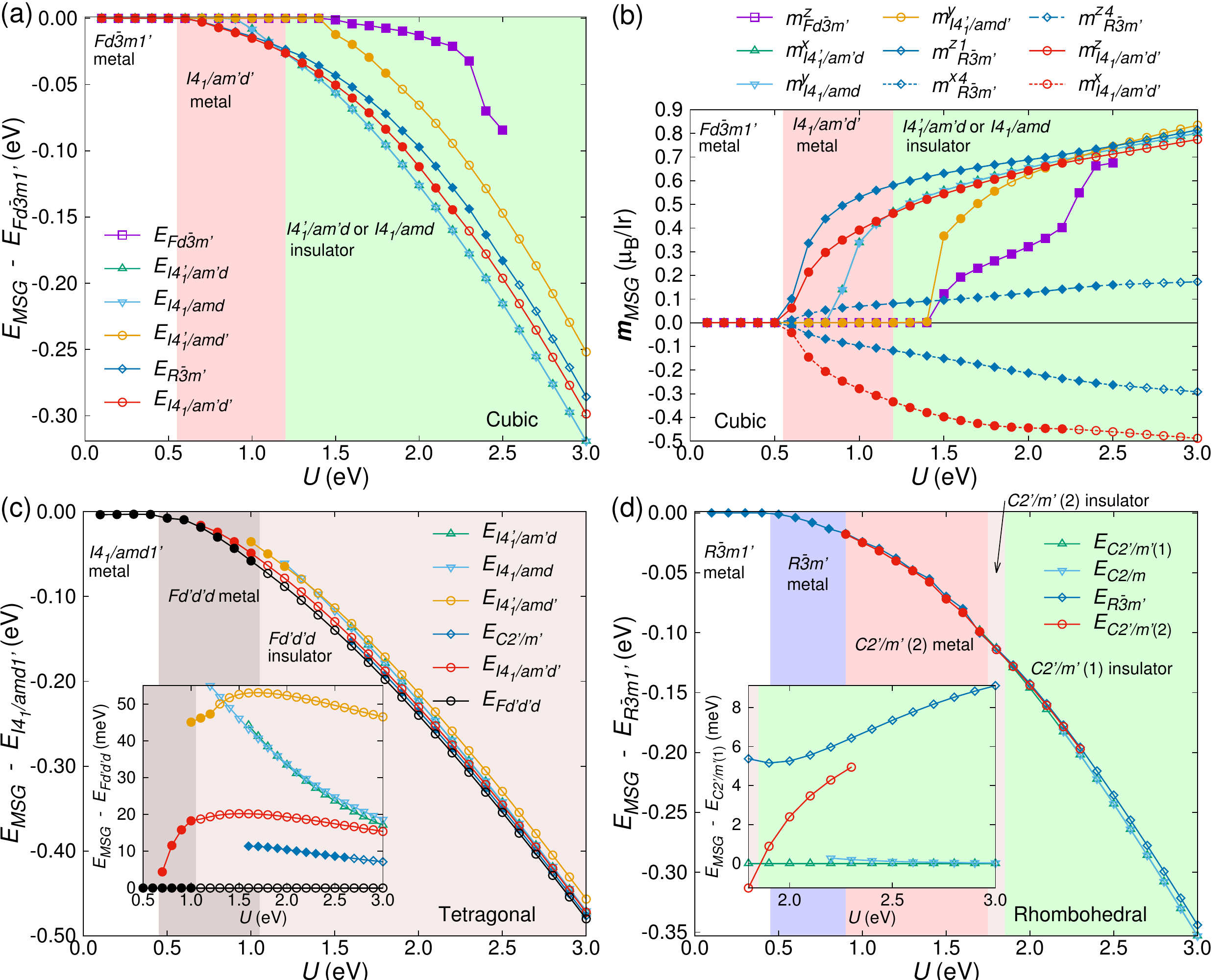}
\caption{(Color online) Phase diagrams of Ir$_2$O$_4$.
  Energies of (semi)stable magnetically ordered states compared to the paramagnetic state as functions of $U$, in the ideal cubic case (a), in the tetragonal case on the MgO(001) substrate (c), and in the rhombohedral case on the LiNbO$_3$(0001) substate (d), respectively. 
  The insets of (c) and (d) highlight the energy differences of semistable states from the ground state in large-$U$ regions.
  (b), Components of the Ir magnetic moments in the magnetically ordered states as functions of $U$ in the cubic case.
  Filled and open symbols stand for (semi)metallic and insulating states. In each case, we have adopted the optimized crystal structure given in Table~1, except the tetragonal or rhombohedral inplane lattice constant determined by the substrate.
}
\label{fig:phasediagram}
\end{center}
\end{figure}
\begin{figure}
  \begin{center}
    \includegraphics[trim = 0 170 0 0, width=8.0cm]{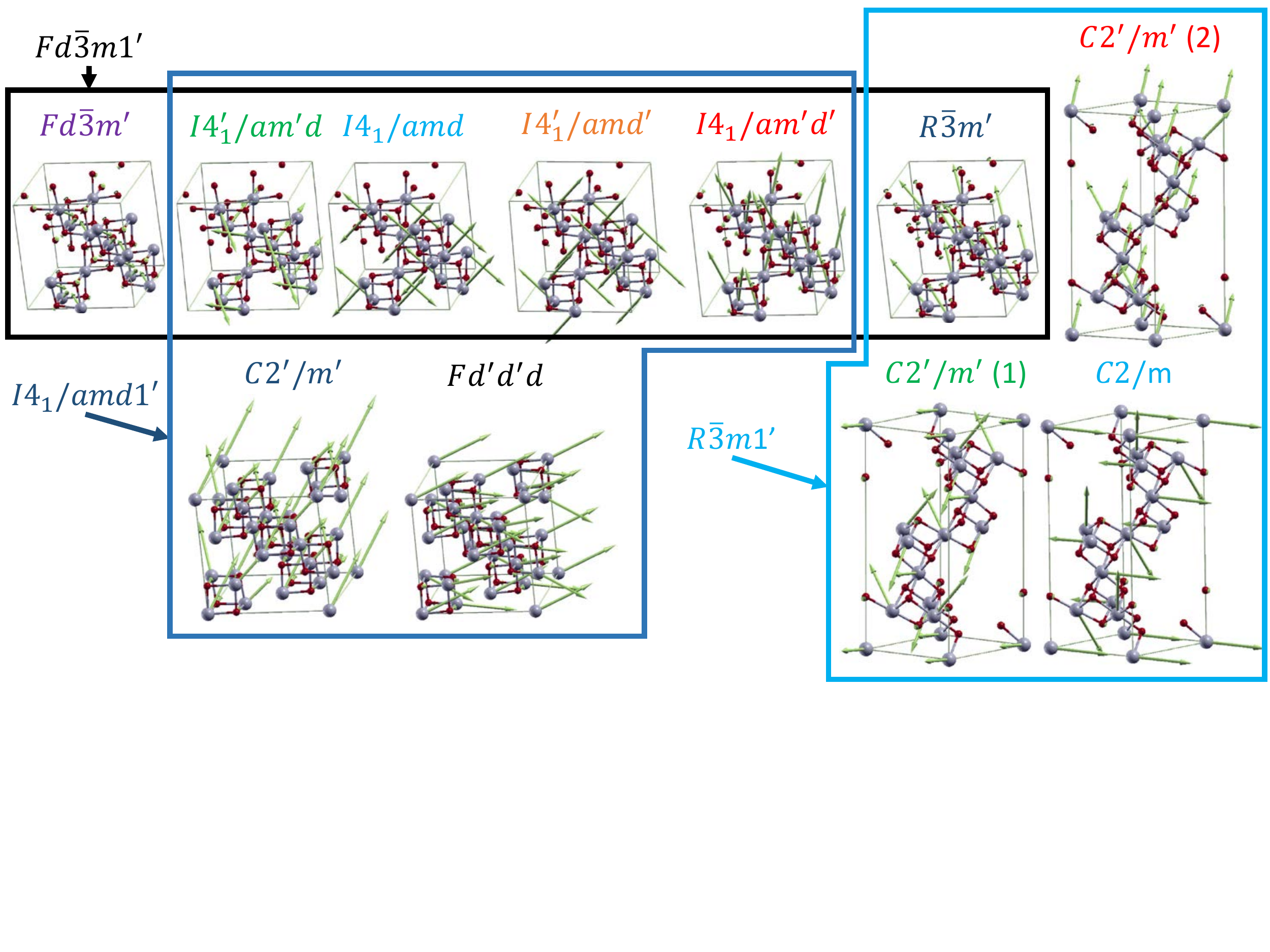}
    \caption{(Color online) Structures and magnetic space groups.
      Magnetically ordered structures assocaited with each magnetic space group are grouped for each of the cubic, tetragonal and rhombohedral case.
    }
  \end{center}
\end{figure}

It has been reported that the tetrahedral Li$_x$Ir$_2$O$_4$ compound can be grown at least on the band-insulating MgO(001) substrate~\cite{kuriyama:10}, which gives a $3.3\%$ compression of the inplane lattice constant $a_{\mathrm{t}}$ from the ideal cubic case. Our first-principles calculations of this tetrahedral Ir$_2$O$_4$ show that the crystal structure (Table~1) belongs to the space group $I4_1/amd$ with a $0.4~\%$ elongation of the out-of-plane lattice constant $c_{\mathrm{t}}$ compared to the cubic case. Figure~2(b) shows the results of LSDA+$U$ calculations on the energies of several (meta)stable states whose magnetic structures are shown in Fig.~3. One can clearly see a phase transition slightly above $U=0.4$~eV from a paramagnetic metal ($I4_1/amd1'$) to a ``2-in, 2-out''-like [100]-ferromagnetic ($Fd'd'd$) metal as in the ideal cubic case, except that the net magnetization is now constrained to be parallel to the substrate. A further increase in $U$ to 1.1~eV drives a metal-insulator transition within $Fd'd'd$. This insulating [100]-ferromagnetic state remains to be the ground state at least up to $U=3.0$~eV, in contrast to the cubic case. The second-lowest-energy solution resembles a ``3-in, 1-out''-like ferromagnet ($C2'/m'$) but with two different pairs of Ir ordered magnetic moment amplitude in the unit cell. This state is obtained for $U\ge1.6$~eV and remain metallic up to $U\le2.6$~eV. The third is a ``2-in, 2-out''-like [001] ferromagnet ($I4_1/am'd'$) with the net magnetization perpendicular to the substrate. Remarkably, the two antiferromagnetic insulating states ($I4_1'/am'd$ and $I4_1/amd$) that form the ground-state manifold in the ideal cubic case shift to much higher energies with a small splitting due to the tetragonality. Thus, it is obvious that the driving force $\delta$ of the $I4_1'/am'd$ and $I4_1/amd$ antiferromagnets has changed across $\delta_c$. Another antiferromagnetic insulating state of $I4'_1/amd'$ driven by the $q$ term also has a high energy as in the cubic case. All these results indicate that this tetragonal Ir$_2$O$_4$ satisfies the first requirement for achieving the U(1) QSLs. Besides, for our choics of $U=1.5$~eV, the rotation angle $\varphi$ of the Ir ordered magnetic moments from the local $\langle111\rangle$ axes is found to be 28.8$^\circ$, which does not change appreciably from 27.3$^circ$ of the insulating $I4_1/am'd'$ ferromagnetic solution obtained for $U=1.5$~eV in the cubic Ir$_2$O$_4$. Therefore, the coupling constant $K$ in the tetragonal Ir$_2$O$_4$ should also be as small as $-0.06$-$0.07$ in the cubic Ir$_2$O$_4$, satisfying the second condition for the stability of the U(1) QSL.

The above results are still not sufficient to judge whether the tetragonal Ir$_2$O$_4$ on the MgO(001) substrate is a ferromagnet of $Fd'd'd$ or the U(1) QSL, since the critical strength of the tetragonal distortion is unknown. The tetragonal distortion splits the spin-ice-rule coupling $J$ into two; $J_\parallel$ and $J_\perp$ for bonds parallel to the substrate and for the others, respectively. Ignoring all the other eight subleading coupling constants allowed by the symmetry, we can make a rough estimate of the difference as $J_\parallel-J_\perp\approx(E_{I4_1/am'd'}-E_{Fd'd'd})/2\sim10$~meV for $U=1.5$~eV. This is not negligible compared with $J\sim30$~meV in the cubic Ir$_2$O$_4$.
Nonetheless, in principle, it is possible to achieve high-temperature quantum spin ice having a U(1) spin liquid behaviour in the tetragonal Ir$_2$O$_4$, if one can take a substrate that makes $a_{\mathrm{t}}$ larger than $\sqrt{2}a_{\mathrm{MgO}}=5.955$~{\AA}
but well below the ideal cubic case $a_{\mathrm{c}}/\sqrt{2}=6.160$~\AA.

Our calculations also show that unlike the tetragonal case, the experimentally realized rhombohedral Ir$_2$O$_4$ on the LiNbO$_3$(0001) substrate~\cite{kuriyama:10} is not advantageous for achieving the U(1) QSL, since insulating ``2-in, 2-out''-like states are not stable except in a narrow region around the metal-insulator transition at $U\sim1.8$~eV, as shown in Fig.~2(d).

The current study highlights the $A$-site deintercalated spinels Ir$_2$O$_4$ with a tetragonal distortion controlled through the substrate as a novel route to a U(1) QSL with a rather high temperature scale given by the spin-ice-rule coupling of $J\sim30$~meV. Then, QSI monopoles should show a coherent quantum dynamics below $J/4\sim87$~K in the analogous QED. This opens a new research direction towards a control of a conserved longitudinal spin current carried by QSI monopoles, which allows for unprecendented potential applications based on QSI monopoles in a magnetic analogue of electronics.

\begin{acknowledgments}
  The work was partially supported by Grants-in-Aid for Scientific Research under Grant No. 24740253, No. 15H03692 and No. 16K05426 from Japan Society for the promotion of Science and under Grant No. 15H01025 from the Ministry of Education, Culture, Sports, and Technology of Japan  and by the RIKEN iTHES project.
Numerical computationss were performed by using the HOKUSAI-Great Wave supercomputing system at RIKEN and supercomputers at ISSP, University of Tokyo and at RIIT, Kyushu University.
\end{acknowledgments}

\bibliographystyle{apsrev}
\bibliography{227}

\begin{thebibliography}{30}
\expandafter\ifx\csname natexlab\endcsname\relax\def\natexlab#1{#1}\fi
\expandafter\ifx\csname bibnamefont\endcsname\relax
  \def\bibnamefont#1{#1}\fi
\expandafter\ifx\csname bibfnamefont\endcsname\relax
  \def\bibfnamefont#1{#1}\fi
\expandafter\ifx\csname citenamefont\endcsname\relax
  \def\citenamefont#1{#1}\fi
\expandafter\ifx\csname url\endcsname\relax
  \def\url#1{\texttt{#1}}\fi
\expandafter\ifx\csname urlprefix\endcsname\relax\def\urlprefix{URL }\fi
\providecommand{\bibinfo}[2]{#2}
\providecommand{\eprint}[2][]{\url{#2}}

\bibitem[{\citenamefont{Lee}(2008)}]{lee:08}
\bibinfo{author}{\bibfnamefont{P.~A.} \bibnamefont{Lee}},
  \bibinfo{journal}{Science} \textbf{\bibinfo{volume}{321}},
  \bibinfo{pages}{1306} (\bibinfo{year}{2008}).

\bibitem[{\citenamefont{Balents}(2010)}]{balents:10}
\bibinfo{author}{\bibfnamefont{L.}~\bibnamefont{Balents}},
  \bibinfo{journal}{Nature} \textbf{\bibinfo{volume}{464}},
  \bibinfo{pages}{199} (\bibinfo{year}{2010}).

\bibitem[{\citenamefont{Molavian et~al.}(2007)\citenamefont{Molavian, Gingras,
  and Canals}}]{molavian:07}
\bibinfo{author}{\bibfnamefont{H.~R.} \bibnamefont{Molavian}},
  \bibinfo{author}{\bibfnamefont{M.~J.~P.} \bibnamefont{Gingras}},
  \bibnamefont{and} \bibinfo{author}{\bibfnamefont{B.}~\bibnamefont{Canals}},
  \bibinfo{journal}{Phys. Rev. Lett.} \textbf{\bibinfo{volume}{98}},
  \bibinfo{pages}{157204} (\bibinfo{year}{2007}).

\bibitem[{\citenamefont{Onoda and Tanaka}(2010)}]{onoda:09}
\bibinfo{author}{\bibfnamefont{S.}~\bibnamefont{Onoda}} \bibnamefont{and}
  \bibinfo{author}{\bibfnamefont{Y.}~\bibnamefont{Tanaka}},
  \bibinfo{journal}{Phys. Rev. Lett.} \textbf{\bibinfo{volume}{105}},
  \bibinfo{pages}{047201} (\bibinfo{year}{2010}).

\bibitem[{\citenamefont{Hermele et~al.}(2004)\citenamefont{Hermele, Fisher, and
  Balents}}]{hermele:04}
\bibinfo{author}{\bibfnamefont{M.}~\bibnamefont{Hermele}},
  \bibinfo{author}{\bibfnamefont{M.~P.~A.} \bibnamefont{Fisher}},
  \bibnamefont{and} \bibinfo{author}{\bibfnamefont{L.}~\bibnamefont{Balents}},
  \bibinfo{journal}{Phys. Rev. B} \textbf{\bibinfo{volume}{69}},
  \bibinfo{pages}{064404} (\bibinfo{year}{2004}).

\bibitem[{\citenamefont{Moessner and Ramirez}(2006)}]{moessner:06}
\bibinfo{author}{\bibfnamefont{R.}~\bibnamefont{Moessner}} \bibnamefont{and}
  \bibinfo{author}{\bibfnamefont{A.}~\bibnamefont{Ramirez}},
  \bibinfo{journal}{Phys. Today} \textbf{\bibinfo{volume}{59}},
  \bibinfo{pages}{24} (\bibinfo{year}{2006}).

\bibitem[{\citenamefont{Gardner et~al.}(2010)\citenamefont{Gardner, Gingras,
  and Greedan}}]{gardner:10}
\bibinfo{author}{\bibfnamefont{J.~S.} \bibnamefont{Gardner}},
  \bibinfo{author}{\bibfnamefont{M.~J.~P.} \bibnamefont{Gingras}},
  \bibnamefont{and} \bibinfo{author}{\bibfnamefont{J.~E.}
  \bibnamefont{Greedan}}, \bibinfo{journal}{Reviews of Modern Physics}
  \textbf{\bibinfo{volume}{82}}, \bibinfo{pages}{53} (\bibinfo{year}{2010}).

\bibitem[{\citenamefont{Chang et~al.}(2014)\citenamefont{Chang, Lees, Watanabe,
  Hillier, Yasui, and Onoda}}]{chang:13}
\bibinfo{author}{\bibfnamefont{L.-J.} \bibnamefont{Chang}},
  \bibinfo{author}{\bibfnamefont{M.~R.} \bibnamefont{Lees}},
  \bibinfo{author}{\bibfnamefont{I.}~\bibnamefont{Watanabe}},
  \bibinfo{author}{\bibfnamefont{A.~D.} \bibnamefont{Hillier}},
  \bibinfo{author}{\bibfnamefont{Y.}~\bibnamefont{Yasui}}, \bibnamefont{and}
  \bibinfo{author}{\bibfnamefont{S.}~\bibnamefont{Onoda}},
  \bibinfo{journal}{Phys. Rev. B} \textbf{\bibinfo{volume}{89}},
  \bibinfo{pages}{184416} (\bibinfo{year}{2014}).

\bibitem[{\citenamefont{Ross et~al.}(2011)\citenamefont{Ross, Savary, Gaulin,
  and Balents}}]{ross:11}
\bibinfo{author}{\bibfnamefont{K.~A.} \bibnamefont{Ross}},
  \bibinfo{author}{\bibfnamefont{L.}~\bibnamefont{Savary}},
  \bibinfo{author}{\bibfnamefont{B.~D.} \bibnamefont{Gaulin}},
  \bibnamefont{and} \bibinfo{author}{\bibfnamefont{L.}~\bibnamefont{Balents}},
  \bibinfo{journal}{Phys. Rev. X} \textbf{\bibinfo{volume}{1}},
  \bibinfo{pages}{021002} (\bibinfo{year}{2011}).

\bibitem[{\citenamefont{Takatsu et~al.}(2016)\citenamefont{Takatsu, Onoda,
  Taniguchi, Kittaka, Kasahara, Kono, Sakakibara, Kato, F{\aa}k, Ollivier
  et~al.}}]{takatsu:15}
\bibinfo{author}{\bibfnamefont{H.}~\bibnamefont{Takatsu}},
  \bibinfo{author}{\bibfnamefont{S.}~\bibnamefont{Onoda}},
  \bibinfo{author}{\bibfnamefont{T.}~\bibnamefont{Taniguchi}},
  \bibinfo{author}{\bibfnamefont{S.}~\bibnamefont{Kittaka}},
  \bibinfo{author}{\bibfnamefont{A.}~\bibnamefont{Kasahara}},
  \bibinfo{author}{\bibfnamefont{Y.}~\bibnamefont{Kono}},
  \bibinfo{author}{\bibfnamefont{T.}~\bibnamefont{Sakakibara}},
  \bibinfo{author}{\bibfnamefont{Y.}~\bibnamefont{Kato}},
  \bibinfo{author}{\bibfnamefont{B.}~\bibnamefont{F{\aa}k}},
  \bibinfo{author}{\bibfnamefont{J.}~\bibnamefont{Ollivier}},
  \bibnamefont{et~al.}, \bibinfo{journal}{Phys. Rev. Lett.}
  \textbf{\bibinfo{volume}{116}}, \bibinfo{pages}{217201}
  (\bibinfo{year}{2016}).

\bibitem[{\citenamefont{Kimura et~al.}(2013)\citenamefont{Kimura, Nakatsuji,
  Wen, Broholm, Stone, Nishibori, and Sawa}}]{kimura:13}
\bibinfo{author}{\bibfnamefont{K.}~\bibnamefont{Kimura}},
  \bibinfo{author}{\bibfnamefont{S.}~\bibnamefont{Nakatsuji}},
  \bibinfo{author}{\bibfnamefont{J.}~\bibnamefont{Wen}},
  \bibinfo{author}{\bibfnamefont{C.}~\bibnamefont{Broholm}},
  \bibinfo{author}{\bibfnamefont{M.}~\bibnamefont{Stone}},
  \bibinfo{author}{\bibfnamefont{E.}~\bibnamefont{Nishibori}},
  \bibnamefont{and} \bibinfo{author}{\bibfnamefont{H.}~\bibnamefont{Sawa}},
  \bibinfo{journal}{Nature communications} \textbf{\bibinfo{volume}{4}},
  \bibinfo{pages}{1934} (\bibinfo{year}{2013}).

\bibitem[{\citenamefont{Harris et~al.}(1997)\citenamefont{Harris, Bramwell,
  McMorrow, Zeiske, and Godfrey}}]{harris:97}
\bibinfo{author}{\bibfnamefont{M.~J.} \bibnamefont{Harris}},
  \bibinfo{author}{\bibfnamefont{S.~T.} \bibnamefont{Bramwell}},
  \bibinfo{author}{\bibfnamefont{D.~F.} \bibnamefont{McMorrow}},
  \bibinfo{author}{\bibfnamefont{T.}~\bibnamefont{Zeiske}}, \bibnamefont{and}
  \bibinfo{author}{\bibfnamefont{K.~W.} \bibnamefont{Godfrey}},
  \bibinfo{journal}{Phys. Rev. Lett.} \textbf{\bibinfo{volume}{79}},
  \bibinfo{pages}{2554} (\bibinfo{year}{1997}).

\bibitem[{\citenamefont{Bramwell and Gingras}(2001)}]{bramwell:01}
\bibinfo{author}{\bibfnamefont{S.~T.} \bibnamefont{Bramwell}} \bibnamefont{and}
  \bibinfo{author}{\bibfnamefont{M.~J.~P.} \bibnamefont{Gingras}},
  \bibinfo{journal}{Science} \textbf{\bibinfo{volume}{294}},
  \bibinfo{pages}{1495} (\bibinfo{year}{2001}).

\bibitem[{\citenamefont{Kim et~al.}(2009)\citenamefont{Kim, Ohsumi, Komesu,
  Sakai, Morita, Takagi, and Arima}}]{kim:09}
\bibinfo{author}{\bibfnamefont{B.~J.} \bibnamefont{Kim}},
  \bibinfo{author}{\bibfnamefont{H.}~\bibnamefont{Ohsumi}},
  \bibinfo{author}{\bibfnamefont{T.}~\bibnamefont{Komesu}},
  \bibinfo{author}{\bibfnamefont{S.}~\bibnamefont{Sakai}},
  \bibinfo{author}{\bibfnamefont{T.}~\bibnamefont{Morita}},
  \bibinfo{author}{\bibfnamefont{H.}~\bibnamefont{Takagi}}, \bibnamefont{and}
  \bibinfo{author}{\bibfnamefont{T.-h.} \bibnamefont{Arima}},
  \bibinfo{journal}{Science} \textbf{\bibinfo{volume}{323}},
  \bibinfo{pages}{1329} (\bibinfo{year}{2009}).

\bibitem[{\citenamefont{Wan et~al.}(2011)\citenamefont{Wan, Turner, Vishwanath,
  and Savrasov}}]{wan:11}
\bibinfo{author}{\bibfnamefont{X.}~\bibnamefont{Wan}},
  \bibinfo{author}{\bibfnamefont{A.~M.} \bibnamefont{Turner}},
  \bibinfo{author}{\bibfnamefont{A.}~\bibnamefont{Vishwanath}},
  \bibnamefont{and} \bibinfo{author}{\bibfnamefont{S.~Y.}
  \bibnamefont{Savrasov}}, \bibinfo{journal}{Phys. Rev. B}
  \textbf{\bibinfo{volume}{83}}, \bibinfo{pages}{205101}
  (\bibinfo{year}{2011}).

\bibitem[{\citenamefont{Ishii et~al.}(2015)\citenamefont{Ishii, Mizuta, Kato,
  Ozaki, Weng, and Onoda}}]{ishii:15}
\bibinfo{author}{\bibfnamefont{F.}~\bibnamefont{Ishii}},
  \bibinfo{author}{\bibfnamefont{Y.~P.} \bibnamefont{Mizuta}},
  \bibinfo{author}{\bibfnamefont{T.}~\bibnamefont{Kato}},
  \bibinfo{author}{\bibfnamefont{T.}~\bibnamefont{Ozaki}},
  \bibinfo{author}{\bibfnamefont{H.}~\bibnamefont{Weng}}, \bibnamefont{and}
  \bibinfo{author}{\bibfnamefont{S.}~\bibnamefont{Onoda}}, \bibinfo{journal}{J.
  Phys. Soc. Jpn.} \textbf{\bibinfo{volume}{84}}, \bibinfo{pages}{073703}
  (\bibinfo{year}{2015}).

\bibitem[{\citenamefont{Tomiyasu et~al.}(2012)\citenamefont{Tomiyasu,
  Matsuhira, Iwasa, Watahiki, Takagi, Wakeshima, Hinatsu, Yokoyama, Ohoyama,
  and Yamada}}]{tomiyasu:12}
\bibinfo{author}{\bibfnamefont{K.}~\bibnamefont{Tomiyasu}},
  \bibinfo{author}{\bibfnamefont{K.}~\bibnamefont{Matsuhira}},
  \bibinfo{author}{\bibfnamefont{K.}~\bibnamefont{Iwasa}},
  \bibinfo{author}{\bibfnamefont{M.}~\bibnamefont{Watahiki}},
  \bibinfo{author}{\bibfnamefont{S.}~\bibnamefont{Takagi}},
  \bibinfo{author}{\bibfnamefont{M.}~\bibnamefont{Wakeshima}},
  \bibinfo{author}{\bibfnamefont{Y.}~\bibnamefont{Hinatsu}},
  \bibinfo{author}{\bibfnamefont{M.}~\bibnamefont{Yokoyama}},
  \bibinfo{author}{\bibfnamefont{K.}~\bibnamefont{Ohoyama}}, \bibnamefont{and}
  \bibinfo{author}{\bibfnamefont{K.}~\bibnamefont{Yamada}},
  \bibinfo{journal}{J. Phys. Soc. Jpn.} \textbf{\bibinfo{volume}{81}},
  \bibinfo{pages}{4709} (\bibinfo{year}{2012}).

\bibitem[{\citenamefont{Sagayama et~al.}(2013)\citenamefont{Sagayama, Uematsu,
  Arima, Sugimoto, Ishikawa, O'Farrell, and Nakatsuji}}]{sagayama:13}
\bibinfo{author}{\bibfnamefont{H.}~\bibnamefont{Sagayama}},
  \bibinfo{author}{\bibfnamefont{D.}~\bibnamefont{Uematsu}},
  \bibinfo{author}{\bibfnamefont{T.-h.} \bibnamefont{Arima}},
  \bibinfo{author}{\bibfnamefont{K.}~\bibnamefont{Sugimoto}},
  \bibinfo{author}{\bibfnamefont{J.~J.} \bibnamefont{Ishikawa}},
  \bibinfo{author}{\bibfnamefont{E.}~\bibnamefont{O'Farrell}},
  \bibnamefont{and}
  \bibinfo{author}{\bibfnamefont{S.}~\bibnamefont{Nakatsuji}},
  \bibinfo{journal}{Phys. Rev. B} \textbf{\bibinfo{volume}{87}},
  \bibinfo{pages}{100403} (\bibinfo{year}{2013}).

\bibitem[{\citenamefont{Kuriyama et~al.}(2010)\citenamefont{Kuriyama, Matsuno,
  Niitaka, Uchida, Hashizume, Nakao, Sugimoto, Ohsumi, Takata, and
  Takagi}}]{kuriyama:10}
\bibinfo{author}{\bibfnamefont{H.}~\bibnamefont{Kuriyama}},
  \bibinfo{author}{\bibfnamefont{J.}~\bibnamefont{Matsuno}},
  \bibinfo{author}{\bibfnamefont{S.}~\bibnamefont{Niitaka}},
  \bibinfo{author}{\bibfnamefont{M.}~\bibnamefont{Uchida}},
  \bibinfo{author}{\bibfnamefont{D.}~\bibnamefont{Hashizume}},
  \bibinfo{author}{\bibfnamefont{A.}~\bibnamefont{Nakao}},
  \bibinfo{author}{\bibfnamefont{K.}~\bibnamefont{Sugimoto}},
  \bibinfo{author}{\bibfnamefont{H.}~\bibnamefont{Ohsumi}},
  \bibinfo{author}{\bibfnamefont{M.}~\bibnamefont{Takata}}, \bibnamefont{and}
  \bibinfo{author}{\bibfnamefont{H.}~\bibnamefont{Takagi}},
  \bibinfo{journal}{Appl. Phys. Lett.} \textbf{\bibinfo{volume}{96}},
  \bibinfo{pages}{182103} (\bibinfo{year}{2010}).

\bibitem[{not({\natexlab{a}})}]{note:LocalFrames}
\bibinfo{note}{For a pyrochlore lattice site $\bm{r}_\mu=\bm{R}+\bm{b}_\mu$
  with the FCC lattice vector $\bm{R}$ and the sublattice vector
  $\bm{b}_\mu=-(\sqrt{3}a/8)\bm{z}_\mu$, we take
  $\{\bm{x}_\mu\}=\frac{1}{\sqrt{6}}\{(1,1,-2), (1,-1,2), (-1,1,2),
  (-1,-1,-2)\}$, $\{\bm{y}_\mu\}=\frac{1}{\sqrt{2}}\{(-1,1,0), (-1,-1,0),
  (1,1,0), (1,-1,0)\}$, $\{\bm{z}_\mu\}=\frac{1}{\sqrt{3}}\{(1,1,1), (1,-1,-1),
  (-1,1,-1), (-1,-1,1)\}$. Then, the phase $\phi_{\bm{r},\bm{r}'}$ in
  Eq.~(\ref{eq:H}) is given by $2\pi/3$, $-2\pi/3$ and 0 for the
  nearest-neighbor bonds parallel to the global $yz$, $zx$ and $xy$
  planes~\cite{onoda:10}.}

\bibitem[{not({\natexlab{b}})}]{note:Openmx}
\bibinfo{note}{Open source package for Material eXplorer
  (http://www.openmx-square.org). We have used the same fully relativistic
  pseudopotentials and pseudoatomic orbitals as in the previous calculations
  for Y$_2$Ir$_2$O$_7$ and Pr$_2$Ir$_2$O$_7$~\cite{ishii:15}, where the
  validity was examined to reproduce results of all-electron full-potential
  calculations based on the FLAPW method; three, three and two radial
  wavefunctions of $s$, $p$ and $d$ orbitals, respectively, for Ir with the
  cutoff radius $7.0a_B$, and three, three and one radial wavefunctions of $s$,
  $p$ and $d$ orbitals, respectively, for O with the cuttoff radius $5.0a_B$,
  with the Bohr radius $a_B$. We employed a $(18,18,18)$ uniform k-point mesh
  in the cubic and tetragonal cases and a $(11,11,3)$ mesh in the rhombohedral
  case.}

\bibitem[{\citenamefont{Witczak-Krempa
  et~al.}(2013)\citenamefont{Witczak-Krempa, Go, and Kim}}]{witczak-krempa:13}
\bibinfo{author}{\bibfnamefont{W.}~\bibnamefont{Witczak-Krempa}},
  \bibinfo{author}{\bibfnamefont{A.}~\bibnamefont{Go}}, \bibnamefont{and}
  \bibinfo{author}{\bibfnamefont{Y.}~\bibnamefont{Kim}},
  \bibinfo{journal}{Phys. Rev. B} \textbf{\bibinfo{volume}{87}},
  \bibinfo{pages}{155101} (\bibinfo{year}{2013}).

\bibitem[{\citenamefont{Anisimov et~al.}(1991)\citenamefont{Anisimov, Zaanen,
  and Andersen}}]{anisimov:91}
\bibinfo{author}{\bibfnamefont{V.~I.} \bibnamefont{Anisimov}},
  \bibinfo{author}{\bibfnamefont{J.}~\bibnamefont{Zaanen}}, \bibnamefont{and}
  \bibinfo{author}{\bibfnamefont{O.~K.} \bibnamefont{Andersen}},
  \bibinfo{journal}{Phys. Rev. B} \textbf{\bibinfo{volume}{44}},
  \bibinfo{pages}{943} (\bibinfo{year}{1991}).

\bibitem[{\citenamefont{Champion et~al.}(2003)\citenamefont{Champion, Harris,
  Holdsworth, Wills, Balakrishnan, Bramwell, \u{C}i\v{u}m\'{a}r, Fennell,
  Gardner, Lago et~al.}}]{champion:03}
\bibinfo{author}{\bibfnamefont{J.~D.~M.} \bibnamefont{Champion}},
  \bibinfo{author}{\bibfnamefont{M.~J.} \bibnamefont{Harris}},
  \bibinfo{author}{\bibfnamefont{P.~C.~W.} \bibnamefont{Holdsworth}},
  \bibinfo{author}{\bibfnamefont{A.~S.} \bibnamefont{Wills}},
  \bibinfo{author}{\bibfnamefont{G.}~\bibnamefont{Balakrishnan}},
  \bibinfo{author}{\bibfnamefont{S.~T.} \bibnamefont{Bramwell}},
  \bibinfo{author}{\bibfnamefont{E.}~\bibnamefont{\u{C}i\v{u}m\'{a}r}},
  \bibinfo{author}{\bibfnamefont{T.}~\bibnamefont{Fennell}},
  \bibinfo{author}{\bibfnamefont{J.~S.} \bibnamefont{Gardner}},
  \bibinfo{author}{\bibfnamefont{J.}~\bibnamefont{Lago}}, \bibnamefont{et~al.},
  \bibinfo{journal}{Phys. Rev. B} \textbf{\bibinfo{volume}{68}},
  \bibinfo{pages}{020401(R)} (\bibinfo{year}{2003}).

\bibitem[{\citenamefont{Onoda and Tanaka}(2011)}]{onoda:10}
\bibinfo{author}{\bibfnamefont{S.}~\bibnamefont{Onoda}} \bibnamefont{and}
  \bibinfo{author}{\bibfnamefont{Y.}~\bibnamefont{Tanaka}},
  \bibinfo{journal}{Phys. Rev. B} \textbf{\bibinfo{volume}{83}},
  \bibinfo{pages}{094411} (\bibinfo{year}{2011}).

\bibitem[{\citenamefont{Savary and Balents}(2012)}]{savary:12}
\bibinfo{author}{\bibfnamefont{L.}~\bibnamefont{Savary}} \bibnamefont{and}
  \bibinfo{author}{\bibfnamefont{L.}~\bibnamefont{Balents}},
  \bibinfo{journal}{Phys. Rev. Lett.} \textbf{\bibinfo{volume}{108}},
  \bibinfo{pages}{037202} (\bibinfo{year}{2012}).

\bibitem[{\citenamefont{Lee et~al.}(2012)\citenamefont{Lee, Onoda, and
  Balents}}]{lee:12}
\bibinfo{author}{\bibfnamefont{S.~B.} \bibnamefont{Lee}},
  \bibinfo{author}{\bibfnamefont{S.}~\bibnamefont{Onoda}}, \bibnamefont{and}
  \bibinfo{author}{\bibfnamefont{L.}~\bibnamefont{Balents}},
  \bibinfo{journal}{Phys. Rev. B} \textbf{\bibinfo{volume}{86}},
  \bibinfo{pages}{104412} (\bibinfo{year}{2012}).

\bibitem[{\citenamefont{Banerjee et~al.}(2008)\citenamefont{Banerjee, Isakov,
  Damle, and Kim}}]{banerjee:08}
\bibinfo{author}{\bibfnamefont{A.}~\bibnamefont{Banerjee}},
  \bibinfo{author}{\bibfnamefont{S.~V.} \bibnamefont{Isakov}},
  \bibinfo{author}{\bibfnamefont{K.}~\bibnamefont{Damle}}, \bibnamefont{and}
  \bibinfo{author}{\bibfnamefont{Y.~B.} \bibnamefont{Kim}},
  \bibinfo{journal}{Phys. Rev. Lett.} \textbf{\bibinfo{volume}{100}},
  \bibinfo{pages}{047208} (\bibinfo{year}{2008}).

\bibitem[{\citenamefont{Yasui et~al.}(2003)\citenamefont{Yasui, Soda, Iikubo,
  Ito, Sato, Hamaguchi, Matsushita, Wada, Takeuchi, Aso et~al.}}]{yasui:03}
\bibinfo{author}{\bibfnamefont{Y.}~\bibnamefont{Yasui}},
  \bibinfo{author}{\bibfnamefont{M.}~\bibnamefont{Soda}},
  \bibinfo{author}{\bibfnamefont{S.}~\bibnamefont{Iikubo}},
  \bibinfo{author}{\bibfnamefont{M.}~\bibnamefont{Ito}},
  \bibinfo{author}{\bibfnamefont{M.}~\bibnamefont{Sato}},
  \bibinfo{author}{\bibfnamefont{N.}~\bibnamefont{Hamaguchi}},
  \bibinfo{author}{\bibfnamefont{T.}~\bibnamefont{Matsushita}},
  \bibinfo{author}{\bibfnamefont{N.}~\bibnamefont{Wada}},
  \bibinfo{author}{\bibfnamefont{T.}~\bibnamefont{Takeuchi}},
  \bibinfo{author}{\bibfnamefont{N.}~\bibnamefont{Aso}}, \bibnamefont{et~al.},
  \bibinfo{journal}{J. Phys. Soc. Jpn.} \textbf{\bibinfo{volume}{72}},
  \bibinfo{pages}{3014} (\bibinfo{year}{2003}).

\bibitem[{not({\natexlab{c}})}]{note:ExchangeCouplings}
\bibinfo{note}{We write the first-princicles total energy as
  $E_{MSG}=H[\{\bm{\tau}_{\bm{r}_\mu}=U_{\bm{r}_\mu}\cdot\bm{m}^{s,\mu}_{MSG}/(2m^{s}_{\mathrm{gr}}\}]+E_{\mathrm{charge}}$
  with a common charge contribution $E_{\mathrm{charge}}$ and the Ir spin
  magnetic moment $\bm{m}^{s,\mu}_{MSG}$ of the sublattice $\mu$ for the
  solution labeled by $MSG$, as well as
  $m^{s}_{\mathrm{gr}}=|\bm{m}^{s,\mu}_{I4_1'/am'd}|$ which is
  $\mu$-independent. While conditions for the doubly degenerate ground states
  of $I4_1'/am'd$ and $I4_1/amd$ are equivalent in the mean-field level, the
  stable rotation angle $\varphi^s$ of $\bm{m}^{s,\mu}_{I4_1/am'd'}$ from the
  local $\langle111\rangle$ axis depends on the exchange coupling constants.
  Solving the set of five equations determine
  $(J,\delta,q,K,E_{\mathrm{charge}})$ as a function of $U$. Metallic solutions
  contain contrbutions from charge degrees of freedom around the Fermi surface
  to the energies, and hence they are excluded from the analysis.}

\end{thebibliography}

\end{document}